\documentclass[twocolumn,amsmath,amssymb,10pt,superscriptaddress,a4paper,letterpaper,fleqn]{revtex4-1}
\usepackage{amssymb}
\usepackage{epsfig}
\usepackage{graphicx}
\usepackage{dcolumn}
\usepackage{array}
\usepackage{bm}
\usepackage{fancyheadings}
\usepackage{longtable}
\usepackage{multirow}
\usepackage{float}
\usepackage{afterpage}  
\usepackage{color}

\setlongtables
\usepackage[breaklinks=true,linkbordercolor={1 1 1}]{hyperref}

\begin{document}
\setcounter{page}{1}

\title{
\qquad \\ \qquad \\ \qquad \\  \qquad \\  \qquad \\ \qquad \\ 
MCNP6 Fission Cross Section Calculations \\
at Intermediate and High Energies
}

\author{S.G. Mashnik}
\email[Corresponding author, electronic address:\\ ]{mashnik@lanl.gov}

\author{A.J. Sierk}

\author{R.E. Prael} 
\affiliation{Los Alamos National Laboratory, Los Alamos, NM 87545, USA }

\date{\today} 

\begin{abstract}
{
MCNP6 has been Validated and Verified (V\&V) against 
intermediate- and high-energy fission-cross-section
experimental data. An error in
the calculation of fission cross sections of $^{181}$Ta
and a few nearby target nuclei
by the CEM03.03 event generator in MCNP6 and al
``bug'' in the calculation of fission cross sections with the
GENXS option of MCNP6 while using the LAQGSM03.03 event
generator were detected during our V\&V work.
After fixing both 
problems, we find that 
MCNP6 using  CEM03.03 and LAQGSM03.03 
calculates fission cross sections
in good agreement with available experimental data for
reactions induced by nucleons, pions, and photons
on both subactinide and actinide nuclei 
at incident energies from several tens of MeV to
about 1 TeV.}
\end{abstract}
\maketitle


\lhead{ND 2013 Article $\dots$}
\chead{NUCLEAR DATA SHEETS}
\rhead{A. Author1 \textit{et al.}}
\lfoot{}
\rfoot{}
\renewcommand{\footrulewidth}{0.4pt}

\section{ INTRODUCTION}
MCNP6 \cite{MCNP6}
is used in various applications
involving fission reactions at low, but also at
intermediate and high energies.
It is critical that it describes such reactions
as well as possible, therefore it is often
validated and verified against available experimental data
and calculations by other models
(see, e.g., \cite{MCNP6,EPJP2011} and refrences therein). 
However, for nuclear reactions
involving fission, generally the transport codes are only tested 
for how well they describe the fission-fragment yields,
and the emission of particles from them (spectra and multiplicities,
both ``prompt'' and ``delayed''). 
Calculation of fission cross sections,
$\sigma_f$,
 themselves by the
transport codes is usually not tested, as it is assumed that data 
libraries used 
at energies below 150 MeV must provide reliable  $\sigma_f$,
 as should the event generators used at higher energies.
Our recent
work \cite{fitaf2012}
shows that this assumption is too
optimistic, as some event generators have problems
describing well some fission cross sections. In such cases,
all the characteristics of the corresponding fission reactions
calculated by the transport codes,
like the yields of fission fragments, spectra and multiplicities
of neutrons and other particles, both ``prompt'' and ``delayed''
would also not be reliable. To assess this situation, 
we test how MCNP6 calculates $\sigma_f$
using the Cascade-Exciton Model (CEM) 
and the Los Alamos version of the Quark-Gluon String Model (LAQGSM)
as realized in the  event generators
CEM03.03 and LAQGSM03.03, respectively, for
reactions induced by nucleons, pions, and photons at incident
energies from several tens of MeV to $\sim 1$ TeV. 

A comprehensive description of the CEM03.03 and LAQGSM03.03 event 
generators can be found in Ref. \cite{Trieste08}
 Therefore,
we present only a brief summary of how CEM03.03 and LAQGSM03.03 
calculate fission cross sections, needed to better understand our
results.

\section{$\sigma_f$ CALCULATION BY CEM AND LAQGSM}

CEM03.03 and LAQGSM03.03 calculate the fission cross sections (and the yields 
of fission products and evaporation of particles
from them) using a modification of the Generalized-Evaporation-Model
code GEM2 by Furihata \cite{GEM2}. GEM2, in its turn, is based on
a modification of the older RAL fission-model code by Atchison
(see \cite{RAL} and references therein). 
 For fissioning nuclei with $Z_j \leq 88$,
GEM2 uses the original Atchison calculation of the neutron emission
width $\Gamma_n$ and fission width $\Gamma_f$ to estimate the fission
probability as
\begin{equation}
P_f = {\Gamma_f \over {\Gamma_f+\Gamma_n}}={1 \over {1+\Gamma_n/\Gamma_f}}.
\end{equation}
Atchison uses \cite{RAL} the Weisskopf and Ewing statistical model
\cite{Weisskopf}. All details on the RAL and GEM2 codes
and all formulas used by them
to calculate  $\sigma_f$
can be found in Refs. \cite{GEM2,RAL}. Let us mention here only that 
in the case of subactinide nuclei, 
the main parameter 
that determines the fission cross sections
calculated by GEM2 is the level-density parameter in the
fission channel, $a_f$ (or more exactly, the ratio $a_f/a_n$), where $a_n$
is the level-density parameter for neutron evaporation. 
GEM2 uses for it the follwing approximation:
\begin{equation}
a_f = a_n \Bigl(1.08926 + 0.01098 ( \chi - 31.08551)^2\Bigr),
\end{equation}
with $\chi = Z^2/A$.

\begin{figure}[!h]
\vspace*{2mm}
\includegraphics[width=0.8\columnwidth]{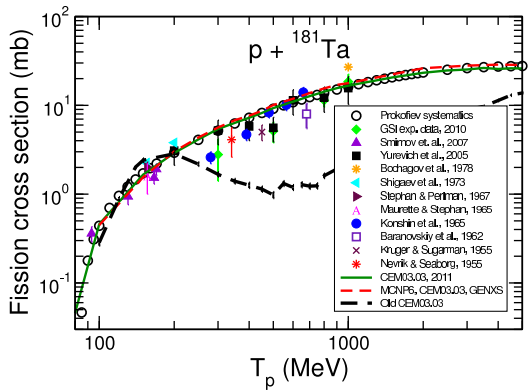}
\caption{Prokofiev systematics \cite{Prokofiev} (open circles) and 
experimental proton-induced fission cross sections of $^{181}$Ta 
(symbols, see detailed references in \cite{fitaf2012}) 
compared with 
our old MCNP6 calculations (black dashed lines) using the CEM03.03
event generator before we fixed the error,
with the corrected CEM03.03 results (green line), 
and with calculations by the updated MCNP6 using the corrected
CEM03.03 event generator (red dashed line), as indicated.}
\label{fig1}
\end{figure}

Furihata obtained
Eq. (2) as the best approximation for the ratio $a_f/a_n$
while calculating  $\sigma_f$ with GEM2 used after the Bertini
IntraNuclear Cascade (INC) model \cite{Bertini},
without taking into account possible preequilibrium emission of 
particles after the INC.
The INC's of CEM and LAQGSM are different from
the Bertini version 
\cite{Bertini}, and both models account for preequilibrium
emission of particles. Hence, 
for any particular reaction,
the mean mass, $<A>$, and charge,
$<Z>$, 
numbers of compound nuclei that may fission, as well as their mean
excitation energy, $<E>$, calculated by CEM and LAQGSM differ from the ones
provided by the Bertini INC. As a result, the best approximation provided by Eq. (2)
for the Bertini INC is not the best for CEM and LAQGSM.
In Ref. \cite{fitaf2003}, we performed
our own fitting of the ratio $a_f/a_n$ 
which works the best for CEM and LAQGSM,
using as ``experimental'' proton-induced $\sigma_f$
the systematics by Prokofiev \cite{Prokofiev}.
After this adjustment for proton-induced reactions, CEM and LAQGSM were 
tested on reactions induced by neutrons, photons, and pions,
and were found to calculate well $\sigma_f$
for these types of reactions 
(see details in \cite{fitaf2003}).

\section{RESULTS}

Kowing from \cite{fitaf2003} 
that both CEM and LAQGSM describe well $\sigma_f$
for various reactions, we did not check 
until very recently how well
MCNP6 and its precursor, MCNPX, using CEM and LAQGSM,
calculated fission cross sections.
The first attempt to calculate  $\sigma_f$
with MCNP6 using CEM03.03
for p + $^{181}$Ta found some
very bad results at proton energies above $\sim 200$ MeV
(see the black dashed line in Fig. 1).
We verified that these MCNP6 results  were not a
problem of the incorporation of CEM03.03 into MCNP6, but were a 
result of an error inadvertently introduced into 
CEM03.03 in 2005 during a major upgrade of the code
\cite{Pavia2006}.

\begin{figure}[!h]
\vspace*{2mm}
\includegraphics[width=0.80\columnwidth,angle=-0]{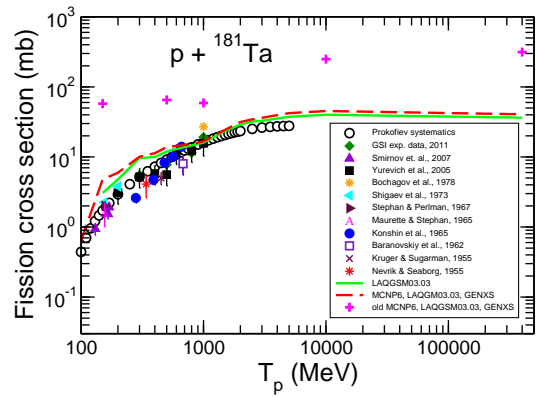}
\caption{Prokofiev systematics \cite{Prokofiev} (open circles) 
and experimental
proton-induced fission cross sections of  $^{181}$Ta 
(symbols, see detailed references in \cite{fitaf2012}) 
compared with 
our results by the updated MCNP6
 using the 
LAQGSM03.03 event generator with the GENXS option 
(red dashed lines) and with calculations by 
LAQGSM03.03 used as a stand alone code
(green lines), as indicated. For comparison,
wrong initial results by an older version of MCNP6
(called ``Beta 1''), before the ``bug'' in the 
calculation of the fission cross section with
the GENXS option while using LAQGSM03.03
was fixed, are shown as well
with several magenta crosses.}
\label{fig2}
\end{figure}

To fix this problem, we have refitted in \cite{fitaf2012} the values 
of  $a_f/a_n$
in GEM2 for $^{181}$Ta. 
Our new CEM03.03 results for $^{181}$Ta are shown
in Fig.\ 1 with a green line.
We have replaced 
in MCNP6 the initial defective CEM03.03 module with the corrected
version with the correct values of  $a_f/a_n$
in GEM2. Results from the updated MCNP6 are
also shown in Fig.\ 1 by a red dashed line.

\begin{figure*}[!htb]
\includegraphics[width=0.7\textwidth]{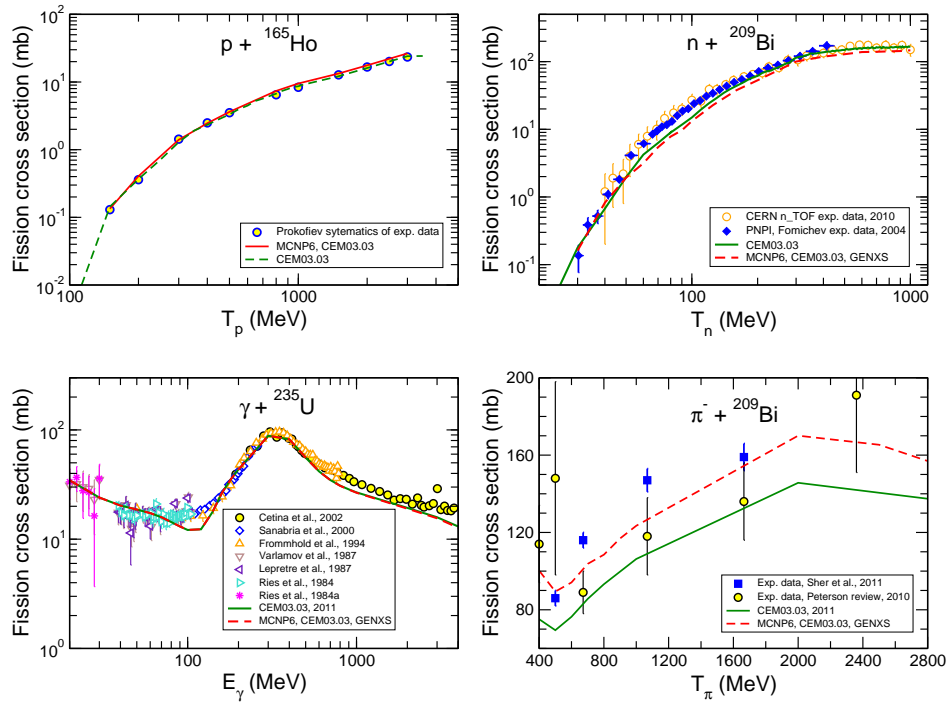}
\caption{Examples of $\sigma_f$ for reactions induced by $p$, $n$,
$\gamma$, and $\pi^-$. Detailed references to measured
data (symbols) can by found in \cite{fitaf2012}. Dfferences in the
absolute values of  $\sigma_f$ calculated with CEM03.03 used as a
stand-alone code and with MCNP6 using CEM are related to different
total reaction cross sections calculated by CEM and used by MCNP6
for these reactions.}
\label{fig3}
\end{figure*}

Having discovered this 2005 error and knowing how it affects
the CEM results, we can understand now why in a recent work 
 \cite{TitarenkoPRC2011}
it was found that CEM03.02 (which has practically the same physics 
as the version CEM03.03 used here) provided such a poor
agreement with the measured yields of the nuclides
produced in proton interactions with $^{181}$Ta and
nearby target nuclei for energies above 250 MeV.  

Unfortunately, 
we also discovered \cite{fitaf2012} that  $\sigma_f$
printed in the output files of MCNP6
while using the GENXS option with LAQGSM03.03 were not
correct; we present an example of such erroneous
results in Fig.\ 2, shown in the plot with magenta crosses.
MCNP6 provides correct results for fission-fragment yields and 
for particle spectra in its output file.
However, an unobserved error in counting the number
of fissioning nuclei in the GENXS portion of MCNP6 when using
the LAQGSM03.03 event generator was present in the initial Beta 1,
version of MCNP6 \cite{MCNP6}. We have fixed that 
error. The current version of MCNP6 
is free of that error and provides in its output
files correct values for $\sigma_f$ (see the red
 dashed line in Fig. 2).

After fixing both these problems, we find \cite{fitaf2012}
that MCNP6 using CEM03.03 and LAQGSM03.03 
calculates $\sigma_f$
in good agreement with available data for
reactions induced 
by $p$, $n$,
$\gamma$, and $\pi^-$ 
on both subactinide and actinide nuclei,
at incident energies from several tens of MeV up to $\sim 1$ TeV
(see an example in Fig. 3 and more details in Ref. \cite{fitaf2012}).

\section{ CONCLUSIONS}

MCNP6 has been validated and verified against 
intermediate- and high-energy fission-cross-section
experimental data. An error in
the calculation of fission cross sections of $^{181}$Ta
and other nearby target nuclei
by the CEM03.03 event generator of MCNP6 and a
``bug'' in the calculation of fission cross sections with the
GENXS option of MCNP6 while using the LAQGSM03.03 event
generator were detected during our current V\&V work.
After fixing both these problems, 
we find that 
MCNP6 using CEM03.03 and LAQGSM03.03 event generators 
calculates  $\sigma_f$
in a good agreement with available experimental data for
reactions induced by nucleons, pions, and photons
on both subactinide and actinide nuclei (from $^{165}$Ho
to $^{239}$Pu) at incident energies from several tens of MeV to
about 1 TeV.

\end{document}